\documentclass[prd,twocolumn,amsmath,amssymb,floatfix,superscriptaddress]{revtex4-1}

\usepackage{graphicx}
\usepackage{amssymb}
\usepackage{amsmath}
\usepackage{hyperref}

\usepackage{color}

\DeclareMathAlphabet\mathbfcal{OMS}{cmsy}{b}{n}
\definecolor{darkgreen}{cmyk}{0.85,0.2,1.00,0.2} 
\definecolor{purple}{cmyk}{0.5,1.0,0,0}

\newcommand{\tX}{{X}}

\def\barray{\begin{array}} 
\def\earray{\end{array}}
\def\be{\begin{equation}}
\def\ee{\end{equation}}
\def\ben{\begin{equation} \nonumber}
\def\een{\end{equation}}
\def\ban{\begin{eqnarray*}}
\def\ean{\end{eqnarray*}}
\def\ba{\begin{eqnarray}}
\def\ea{\end{eqnarray}}

\def\({\left(}
\def\){\right)}

\newcommand{\tr}[1]{[#1]}

\newcommand{\sgisig}{ {\boldsymbol{\gamma}}}
\newcommand{\ul}[3]{#1^{#2}_{\hphantom{#2}#3}}

\newcommand{\fid}{\Sigma}
\newcommand{\bgamma}{\boldsymbol{\gamma}}
\newcommand{\bfid}{\boldsymbol{\Sigma}}
\newcommand{\bg}{{\bf g}}

\newcommand{\stucky}{St\"{u}ckel\-berg}
\newcommand{\af}{a_F}
\newcommand{\daf}{\dot a_F}

\newcommand{\ta}{ a}

\newcommand{\lamfid}{\lambda}
\newcommand{\afrw}{a_F}

\begin{document}

\title{Self-accelerating Massive Gravity: Bimetric Determinant Singularities}
\author{Pierre Gratia}
\email{pgratia@uchicago.edu}
\affiliation{Department of Physics, University of Chicago, Chicago, Illinois 60637, U.S.A.}
\author{Wayne Hu}
\email{whu@background.uchicago.edu}
\affiliation{Kavli Institute for Cosmological Physics, Department of Astronomy \& Astrophysics,  Enrico Fermi Institute, University of Chicago, Chicago, Illinois 60637, U.S.A.}
\author{Mark Wyman}
\email{markwy@oddjob.uchicago.edu}
\affiliation{Kavli Institute for Cosmological Physics, Department of Astronomy \& Astrophysics,  Enrico Fermi Institute, University of Chicago, Chicago, Illinois 60637, U.S.A.}
\affiliation{Center for Cosmology and Particle Physics, Department of Physics, New York University, New York, NY, 10003, U.S.A.}
\begin{abstract}
The existence of two metrics in massive gravity theories in principle allows solutions where there are singularities in new scalar invariants jointly constructed 
from them.    These configurations  occur when the two metrics differ substantially from each other, as in black hole and cosmological solutions.   
{The simplest class of such singularities are determinant singularities. We investigate whether the dynamics of bimetric massive gravity -- where the second metric is allowed to evolve jointly with the spacetime metric -- can avoid these singularities.}
We show that it is still possible to specify non-singular initial conditions that
evolve to a determinant singularity.   Determinant singularities are a feature of massive gravity of both fixed and dynamical metric type.
\end{abstract}

\maketitle
\section{Introduction}

Massive gravity is a theory with two metrics.
In the simplest version, only the usual spacetime
metric  is dynamical; the second metric is taken to be static and typically flat \cite{deRham:2010ik,deRham:2010kj,Hassan:2011hr}.
When the spacetime metric evolves to a point where it deviates far from the second metric, massive gravity enters an interesting regime where singularities in scalar invariants built from the two metrics can arise. By allowing the second metric to evolve with its own dynamics in the so-called bimetric or bigravity theory  \cite{Hassan:2011zd},
it is possible that the character of these singularities can change.

These issues have been explored in detail for black hole solutions \cite{Isham:1971gm,1978PhRvD..18.1047I,Gruzinov:2011mm,Deffayet:2011rh,Volkov:2012wp,Babichev:2013una}.  Indeed, the bimetric theory allows a different
class of solutions from those of the flat metric theory  \cite{Koyama:2011xz},
where the two metrics are simultaneously diagonalizable and horizons coincide.   Being
static solutions, it is however unclear as to whether dynamical systems evolve into
these or other solutions.    

A simpler case in which a singularity arises dynamically was studied for the fixed flat metric case in Ref.~\cite{Gratia:2013gka}.   Here,
the spacetime metric evolves from non-singular initial conditions to a determinant singularity in unitary gauge where the flat metric is in standard Minkowski form. This implies the presence of
a coordinate invariant singularity in the ratio of determinants of the two metrics.   Although the
theory is formally undefined at this point, one can smoothly join solutions on either
side of the singularity with the help of vielbeins, or equivalently \stucky\ fields.

In this {\it Brief Report}, we study the impact of bimetric dynamics on determinant singularities.  We begin in \S \ref{sec:bimetric} with a brief review of the bimetric theory and
continue in \S \ref{sec:isotropic} with the construction of exact isotropic solutions.   
We address the determinant singularity in \S \ref{sec:determinant} and discuss
these results in \S \ref{sec:discussion}.

\section{Bimetric Massive Gravity}
\label{sec:bimetric}

The Boulware-Deser ghost-free bimetric massive gravity Lagrangian is \cite{Hassan:2011zd}
\begin{align}
\label{drgt}
\mathcal{L}_G &=\frac{M_{\rm pl}^2}{2}\sqrt{-g}\left[ R-\frac{m^2}{4} \mathcal{U}\left(\sgisig\right) + \sqrt{\frac{-\fid}{-g}}\frac{\cal R}{\epsilon}\right],
\end{align}
where $R$ is the Ricci scalar for the $\bg$ metric to which matter is coupled and
${\cal R}$ is that of the second metric $\bfid$.   Here
$M_{\rm pl}=(8\pi G)^{-1}$ is the reduced Planck mass and $\epsilon$ allows for the second metric to have a different Planck mass.   The massive gravity potential term $\mathcal{U}$ is constructed from the  square root matrix $\bgamma$
\begin{equation}
\ul{(\bg^{-1}\bfid)}{\mu}{\nu}  \equiv \ul{(\bgamma^2)}{\mu}{\nu} = \ul{\gamma}{\mu}{\alpha} \ul{\gamma}{\alpha}{\nu} 
\end{equation}
such that
\begin{equation}
\frac{\mathcal{U}}{4} = \sum_{k=0}^4 \frac{\beta_k}{k!} F_k,
\end{equation}
where \cite{deRham:2010ik,deRham:2010kj}
\begin{align}
F_0(\bgamma) & = 1, \nonumber\\
F_1(\bgamma) & = \tr{\bgamma}, \nonumber\\
F_2(\bgamma) & =  \tr{\bgamma}^2 - \tr{\bgamma^2} , \\
F_3(\bgamma) & =\tr{\bgamma}^3 - 3 \tr{\bgamma} \tr{\bgamma^2} + 2 \tr{\bgamma^3} , \nonumber\\
F_4(\bgamma) &= \tr{\bgamma}^4 - 6 \tr{\bgamma}^2 \tr{\bgamma^2} + 3 \tr{\bgamma^2}^2 + 8 \tr{\bgamma} \tr{\bgamma^3}  
- 6 \tr{\bgamma^4} ,
\nonumber
\end{align}
and $[\,]$ denotes the trace of the enclosed matrix.  
 To avoid confusion,
we refrain from raising and lowering indices where possible; hence $\bg^{-1}$ rather than 
$g^{\mu\nu}$. 

The bimetric theory is parameterized by
 a graviton mass $m$, the ratio of squared Planck masses $\epsilon$ and 
\begin{align}
\beta_0 &= -12 (1+ 2\alpha_3+2\alpha_4), \nonumber\\
\beta_1 &= 6(1 + 3 \alpha_3 + 4\alpha_4),\nonumber\\
\beta_2 &= -2(1+ 6 \alpha_3+12\alpha_4 ), \\
\beta_3 &= 6(\alpha_3+ 4\alpha_4), \nonumber\\
\beta_4 &= -24 \alpha_4,\nonumber
\end{align} 
or equivalently $\{\alpha_3,\alpha_4\}$. 
Varying the action with respect to each of the metrics gives two Einstein equations
\begin{align}
{R}^{\mu}_{\hphantom\mu\nu}-\frac{1}{2}{R}\delta^{\mu}_{\hphantom\mu\nu}&=  m^2 {T}^{\mu}_{\hphantom\mu\nu} +  \frac{1}{M_{\rm pl}^2}{T^{(\rm m)}}{\vphantom{T}}^{\mu}_{\hphantom\mu\nu} ,\nonumber\\
\mathcal{R}^{\mu}_{\hphantom\mu\nu}-\frac{1}{2}\mathcal{R}\delta^{\mu}_{\hphantom\mu\nu}&= \epsilon m^2 \mathcal{T}^{\mu}_{\hphantom\mu\nu} ,
\label{eqn:Einstein}
\end{align}
where the potential term supplies an effective stress energy 
for both metrics, whereas the matter stress energy $T^{(m)}$ is coupled only to $\bg$.  The construction of ${T}^{\mu}_{\hphantom\mu\nu}$ out of $\sgisig$ is given in Eq.~(7) of Ref.~\cite{Gratia:2012wt} and the stress-tensor source for
$\bfid$ is given by \cite{Volkov:2013roa}
\begin{eqnarray}
{\cal T}^\mu_{\hphantom{\nu}\nu} &=&- \sqrt{ \frac{-g}{-{\fid}} } \left[T^\mu_{\hphantom{\nu}\nu}+
 \frac{{\cal U}}{8} \delta^\mu_{\hphantom{\nu}\nu} \right].
\label{eqn:stress2}
\end{eqnarray}
Interestingly, this relation between the stress-tensors involves the ratio of metric determinants, which can become singular.
Nonetheless,
as we shall see next, for self-accelerating solutions both stress tensors are simply 
constants given by the parameters of the theory.

\section{Exact Bi-Isotropic Solutions}
\label{sec:isotropic}

Exact self-accelerating solutions of bimetric massive gravity can be constructed
when the two metrics are simultaneously isotropic 
\begin{align}
g_{\mu\nu}dx^{\mu}dx^{\nu} &=-b^2(r,t) dt^2+a^2(r,t)(dr^2+r^2d\Omega^2), \\
f_{ab} dx^a dx^b & = -\beta^2(g,f) df^2+\alpha^2(g,f)(dg^2+g^2d\Omega^2),\nonumber
\label{eqn:metric}
\end{align}
where $f_{ab}$ is the representation of $\bfid$ in the so-called unitary gauge and
 $f(r,t)$ and $g(r,t)$, not to be confused with the determinants of the respective metrics, give the transformation between this coordinate system and the one
 used for $\bg$.   Note that they
represent  auxiliary \stucky\ fields 
\begin{equation}
\phi^0 = f(t,r), \quad \phi^i = g(t,r) \frac{x^i}{r},
\end{equation}
such that the second metric in the same coordinate system as $\bg$ is 
\begin{equation}
\fid_{\mu\nu} = \partial_\mu \phi^a \partial_\nu \phi^b f_{ab}.
\end{equation}
  In general, the number of gravitational degrees of freedom is 7 -- the 5 polarization states of one massive graviton,
together with 2 polarizations of one massless graviton. 
In this representation, the extra polarization states of massive gravity are 
carried by the \stucky\ fields.  While there are 4 \stucky\ fields, the ghost-free construction eliminates one degree of freedom and the assumption of bi-isotropy eliminates two more \cite{Wyman:2012iw} leaving the pair
of \stucky\ fields as a single degree of freedom on top of the two usual tensor degrees of
freedom of the two metrics.   

 Bi-isotropy allows us to express the potential as \cite{Gratia:2012wt,Motohashi:2012jd}
\begin{equation}
{\frac{\mathcal U}{4}} = P_0\left( \frac{\alpha g}{\ta r} \right) + \sqrt{\tX}P_1\left( \frac{\alpha g}{\ta r} \right)
+  W P_2\left( \frac{\alpha g}{\ta r} \right),
\label{eqn:genpot}
\end{equation}
where the $P_n$ polynomials are
\begin{align}
P_0(x) &= - 12 - 2 x(x-6) - 12(x-1)(x-2)\alpha_3 
\nonumber\\&\qquad -24(x-1)^2\alpha_4 ,\nonumber\\
P_1(x) &= 2 (3 -2 x)  +  6(x-1)(x-3)\alpha_3 +   24(x-1)^2 \alpha_4,\nonumber\\
P_2(x) &= -2 + 12 (x-1) \alpha_3 - 24(x-1)^2 \alpha_4.
\end{align}
Here
\begin{eqnarray}
X(r,t) &=& \left( \frac{\beta}{b} \dot f + \mu \frac{\alpha}{a} g'\right)^2 -
                \left( \frac{\alpha}{b} \dot g + \mu \frac{\beta}{a} f'\right)^2, \nonumber\\
W(r,t) &=& \mu \frac{\alpha\beta}{ab} \left(\dot f g' - \dot g f' \right),
\label{eqn:XW}
\end{eqnarray}
are related to the $t-r$ block of $\sgisig$ as $\sqrt{X} = [\sgisig_2]$ and $W= {\rm det}\sgisig_2$
whereas
\begin{equation}
\sgisig^2 = \left(
      \begin{array}{cccc}
        \dfrac{\beta^2\dot{f}^2-\alpha^2\dot{g}^2}{b^2} & \dfrac{\beta^2\dot{f}f'-\alpha^2\dot{g}g'}{b^2} & 0 & 0 \\
        \dfrac{\alpha^2 \dot{g}g'-\beta^2 \dot{f}f'}{\ta ^2} & \dfrac{-\beta^2 f'^2+\alpha^2 g'^2}{\ta ^2} & 0 & 0\\
        0 & 0 & 
        \dfrac{\alpha^2 g^2}{\ta ^2r^2} & 0\\
        0 & 0 & 0 & \dfrac{\alpha^2 g^2}{\ta ^2r^2}
      \end{array} \right). \nonumber
  \end{equation}
  
 The  branch choice in the solution to the matrix square root of $\sgisig^2$ specifies 
$\mu =\pm 1$ which remains constant even if $W$ changes sign 
\cite{Gratia:2013gka} (cf.~\cite{D'Amico:2011jj}).
Varying the action with respect to the \stucky\ fields gives the equations of motion for
$f$ and $g$.    For any bi-isotropic pair of metrics, these equations are exactly solved by
$P_1(x_0)=0$
yielding
\begin{equation}
x_0 = \frac{ 1 + 6\alpha_3 + 12\alpha_4 \pm \sqrt{ 1+ 3\alpha_3 + 9\alpha_3^2 - 12 \alpha_4}}{3 (\alpha_3+4\alpha_4)}, \label{gsol}
\end{equation}
and
\begin{equation}
\frac{\alpha g}{ar} = x_0 .
\label{eqn:geqn}
\end{equation}
Note that as $\alpha_3 \rightarrow -4\alpha_4$ one branch of Eq.~(\ref{gsol}) remains finite.
On both, this consistency condition (\ref{eqn:geqn}) for self-accelerating solutions requires that the respective radial coordinates are algebraically related.  

The stress-energy source for the $\bg$ metric is then a cosmological constant \cite{Gratia:2012wt,Motohashi:2012jd}
\begin{eqnarray}
T^\mu_{\hphantom{\nu}\nu} = -\frac{1}{2} P_0(x_0) \delta^\mu_{\hphantom{\nu}\nu}.
\end{eqnarray}
Since this relation holds for any isotropic metric, the interaction potential term acts as 
a cosmological constant for any isotropic distribution of matter, not just vacuum or homogeneous ones.

Moreover, since 
\begin{equation}
\sqrt{ \frac{-\Sigma}{-g} }  = {\rm det}{\sgisig} = x_0^2 W,
\end{equation} 
the stress tensor source to the second metric
\begin{align}
{\cal T}^\mu_{\hphantom{\nu}\nu} &= -\sqrt{ \frac{-g}{-\Sigma} }  W \frac{P_2(x_0)}{2}\nonumber\\
& =-\frac{1}{x_0^2}   \frac{P_2(x_0)}{2}   \delta^\mu_{\hphantom{\nu}\nu}
\label{eqn:Tsecond}
\end{align} 
is also a constant \cite{Volkov:2012cf,Volkov:2012zb,Volkov:2013roa}.   
Note that the stress tensor remains constant even through a determinant singularity where
$\sqrt{{g}/{\Sigma}} \rightarrow \infty$. Given the identity
\begin{equation}
\frac{1}{2} P_0(x) + \frac{1}{2} P_2(x)  + P_1(x) + (x-1)^2=0
\end{equation}
and $P_1(x_0)=0$, if one metric has a positive cosmological constant, the other has a 
negative one \cite{Volkov:2013roa} but both metrics may have a negative cosmological constant.  

Since there is no matter source to $\bfid$, the second Einstein equation (\ref{eqn:Einstein}) is then solved by a 
de Sitter metric in isotropic coordinates
\begin{align}
\alpha(g) &=\frac{1}{1+ \lamfid (g/x_0)^2/4}, \nonumber\\
\beta(g) &= \frac{1- \lamfid (g/x_0)^2/4}{1+ \lamfid (g/x_0)^2/4},
\label{eqn:dSI}
\end{align}
 where
 \begin{equation}
 \lamfid =\frac{ \epsilon m^2}{6}  P_2(x_0).
 \end{equation}
 Of course, as $\epsilon \rightarrow 0$, so does $\lambda$, and the second metric takes the Minkowski form of the
 original massive gravity theory in unitary gauge \cite{deRham:2010ik,deRham:2010kj}.

Note that these results are independent of the solution for $f$ which relates unitary or 
isotropic $\bfid$-time
to  isotropic $\bg$-time.
There are in fact many  solutions for this relation 
that give the same stress tensor and metric
structure individually.    They are specified by solving the second equation of motion
\cite{Wyman:2012iw,Gratia:2013gka}
\begin{equation}
P_1' (x_0 + \frac{W}{x_0} - \sqrt{X})=0.
\label{eqn:feqn}
\end{equation} 
Aside from the special parameter choice of $P_1'(x_0)=0$, where
$12\alpha_4=1+3\alpha_3+9\alpha_3^2$, this equation governs
the evolution of $f$.
Importantly, it remains non-singular as the determinant $W$ goes to zero.  
For the special parameter choice, more static solutions exist
\cite{Nieuwenhuizen:2011sq} but the initial value problem 
in $f,g$ is then ill-posed \cite{Gratia:2013gka}.

Using Eq.~(\ref{eqn:XW}) and (\ref{eqn:geqn}), we can see that Eq.~(\ref{eqn:feqn}) is a nonlinear partial differential equation for $f$ whose solutions are specified by boundary conditions such as
$f(0,t)$ \cite{Wyman:2012iw}.
Note that for a fixed $\lamfid$, both $f$ and $g \propto x_0$ and so a solution for a single
set of massive gravity parameters $\alpha_3$, $\alpha_4$ but arbitrary $\lamfid$ can be scaled to any 
choice \cite{Wyman:2012iw}.
The determinant singularity we discuss next is related to a specific choice of $f(0,t)$ in the solution to
Eq.~(\ref{eqn:feqn}).

\section{Determinant Singularity}
\label{sec:determinant}

Given that metric determinants appear in the Einstein equations (\ref{eqn:Einstein}) through (\ref{eqn:stress2}), it is interesting to examine whether the nature of determinant
singularities in fixed metric massive gravity changes when the second metric becomes
dynamical.  One might expect that a singularity that impacts the equations of motion would
be dynamically avoided.   We shall see that none of them exhibit singular behavior at a
determinant singularity.

In the fixed flat metric theory, we can easily construct solutions that evolve from non-singular
initial conditions to a determinant singularity.   By a coordinate transformation, this singularity
can be hidden from  either metric individually but not both simultaneously. 
The simplest example is that of  an open FRW universe in the $\bg$ metric 
\cite{Gumrukcuoglu:2011ew}
with a negative
cosmological constant term from the interaction potential \cite{Gratia:2013gka}.
 {Here, the singularity
occurs when an initial expansion turns to contraction because of the presence of negative stress-energy}.

Now let us consider how the dynamics of the second metric alter this singular solution.
The open FRW spacetime metric in isotropic coordinates is given by
\begin{equation}
ds^2 = -dt^2 + \left[ \frac{\af(t)}{1+K r^2/4} \right]^2 (dr^2 + r^2 d\Omega^2),
\end{equation}
where the scale factor $\af$ obeys the ordinary Friedmann equation with
spatial curvature $K<0$ 
\begin{equation}
\left( \frac{\daf}{\af} \right)^2 +\frac{K}{\af^2} = \frac{\rho^{({\rm m)}} }{3M_{\rm pl}^2} + \frac{m^2}{6}P_0(x_0).
\end{equation}
By choosing $\alpha_3$ and $\alpha_4$ appropriately, we can make $P_0<0$ and hence the $\bg$ metric
evolves to a point where $\daf=0$.  

We next  solve Eq.~(\ref{eqn:feqn}) for the relationship between the two time 
coordinates $f$ and $t$.   Transforming the isotropic radial coordinate $r$ to the
 dimensionless angular diameter distance
\begin{equation}
y = \frac{\sqrt{-K}  r}{1+K r^2/4},
\end{equation}
we obtain
\begin{eqnarray}
&&y^2 \left[ 1- (\lamfid/K) \afrw^2 \right] \dot f^2 
- 2y(1+y^2) \frac{\daf}{\af} \dot f \frac{\partial f}{\partial y}\nonumber\\&& \qquad
- \frac{1+y^2}{\af^2} [ K+ y^2(\lamfid \af^2 - \daf^2)]\left(\frac{\partial f}{\partial y}\right)^2\nonumber\\
&&\qquad =  x_0^2 y^2 \frac{ K -\lamfid \af^2 + \daf^2}{K + \lamfid \af^2 y^2}.
\end{eqnarray}
First note that as $\lamfid \rightarrow 0$, we recover the solution for a fixed flat second metric
\cite{Gratia:2013gka}
\begin{equation}
\lim_{\lamfid \rightarrow 0} f \equiv f_0 = {x_0} \af \sqrt{\frac{1+ y^2}{-K}},
\end{equation}
with the boundary condition $f(0,t) \propto \af(t)$.
The determinant singularity appears since both $f_0$ and $g \propto \af$ and thus
$W=0$ when $\daf=0$ by virtue of Eq.~(\ref{eqn:XW}).

Now let us check what happens for $\lamfid \ne 0$.  Since  $\lamfid=0$ and $f=f_0$ 
represents a determinant singularity, the simplest test for whether bimetric dynamics automatically avoids determinant singularities is to solve Eq.~(\ref{eqn:feqn}) perturbatively for 
a finite $\lamfid/K \ll 1$.  Even in this limit, there are many solutions to this equation
corresponding to different choices of the perturbed boundary condition $f(0,t)$.  The 
simplest choice is
\begin{eqnarray}
f(y,t) &=& f_0\Big[ 1 -\frac{1}{6} (-1 + 2 y^2)(\lamfid\af^2 /K) \nonumber\\
&&+\frac{1}{40}(3-4y^2+8y^4) (\lamfid\af^2 /K)^2 +\ldots \Big]
\end{eqnarray}
Since $\alpha(g) g \propto \af$, Eq.~(\ref{eqn:XW}) implies that this solution 
retains a determinant singularity at $\daf=0$.  
Other solutions can alter the time at which the determinant becomes singular as a function of
radius.  Nonetheless a determinant singularity must appear in all solutions since $f$ remains perturbatively close to $f_0$.  $W$ changes sign during the evolution through turnaround and must therefore
pass through zero.  Likewise, since neither the stress source (\ref{eqn:Tsecond}) nor the \stucky\
field equations (\ref{eqn:feqn}) become singular for $W=0$, we expect that the inability of bimetric dynamics
to prevent a determinant singularity is not limited to models with $\epsilon \ll 1$.

\section{Discussion}
\label{sec:discussion}

While the bimetric theory of massive gravity allows the second metric to evolve in response
to the first, it does not automatically resolve issues arising from the very existence of two metrics that may evolve to become very different from each other.    We have explicitly shown here
that it is still possible to construct solutions where a determinant singularity arises from the
evolution of non-singular initial conditions.   

This singularity cannot be removed by a 
coordinate transformation but the non-singular equations of motion imply that solutions can
be matched on either side of the singularity.   The curvature of both metrics remains finite
through the singularity and its presence is hidden from observables in the matter sector.   
Hence the existence of determinant singularities is a peculiar but perhaps not pathological feature of both fixed and dynamical bimetric massive gravity theories.

\smallskip{\em Acknowledgments.--}   We thank Rachel Rosen for discussions and assistance during the start of this project.  
PG was supported by the National Research Fund Luxembourg through grant BFR08-024. WH and MW were supported by Kavli Institute for Cosmological Physics at the University of Chicago through grants NSF PHY-0114422 and NSF PHY-0551142  and an endowment from the Kavli Foundation and its founder Fred Kavli and by the David and Lucile Packard Foundation and by U.S.~Dept.\ of Energy contract DE-FG02-90ER-40560.  WH thanks ICG Portsmouth for hospitality where part of this work was completed.

\vfill
\bibliography{bidet}
\vfill
\end{document}